\documentclass[twocolumn,nofootinbib,prd,a4paper,showpacs,showkeys]{revtex4}
\usepackage{graphicx,amssymb}
\usepackage[intlimits]{amsmath}
\usepackage{natbib}

\begin{document}

\title{Neutrino spectrum from the pair-annihilation process in the hot stellar plasma}

\date{\today}

\author{M. Misiaszek}
\affiliation{M. Smoluchowski Institute of Physics, Jagiellonian University, Reymonta 4, 30-059 Cracow, Poland}
\author{A. Odrzywo\l{}ek}
\affiliation{M. Smoluchowski Institute of Physics, Jagiellonian University, Reymonta 4, 30-059 Cracow, Poland}
\email{odrzywolek@th.if.uj.edu.pl}
\author{M. Kutschera}
\affiliation{M. Smoluchowski Institute of Physics, Jagiellonian University, Reymonta 4, 30-059 Cracow, Poland}
\affiliation{ The Henryk Niewodniczanski Institute of Nuclear Physics, 
Polish Academy of Sciences,
152 Radzikowskiego St,
31-342 Cracow, Poland}

\begin{abstract}
An new method of calculating the energy spectrum of neutrinos 
and antineutrinos produced 
in the electron-positron annihilation processes in hot stellar plasma is 
presented.
Detection of these neutrinos, produced copiously in 
the presupernova 
which is evolutionary advanced neutrino-cooled star, may serve in  future 
as a trigger of  
pre-collapse early warning system. Also, observation of neutrinos will 
probe final stages of thermonuclear burning 
in the presupernova. 

The spectra obtained with the new method are compared to
Monte~Carlo simulations. To achieve high accuracy in the energy range of 
interest, determined by neutrino detector thresholds, 
 differential cross-section for 
production of the antineutrino, 
previously unknown in an explicit form, is calculated as a function of 
energy in the plasma rest frame. 
Neutrino spectrum is obtained as a 3-dimensional integral, computed with 
the use of the Cuhre algorithm of at least 5\% accuracy.  Formulae for 
the mean neutrino energy and its dispersion 
are given as a combination of Fermi-Dirac integrals. Also, useful analytical 
approximations of the whole spectrum are shown.
\end{abstract}
\maketitle

\section{Introduction \& Motivation}

We present here a new method which allows us to calculate the spectrum of 
neutrinos 
produced by thermal pair-annihilation processes in the hot plasma in the core of
massive pre-supernova star. The
paper is organized as follows: In sect.II the electron-positron pair 
annihilation processes into neutrinos in the
Standard Model are briefly summarized. 
Monte~Carlo simulations are discussed in  section~\ref{monte}. Simple
estimates of the  average energy of neutrinos
based on the Dicus cross-section \cite{Dicus} 
are presented in Subsection~\ref{avg}.

The complete spectrum is obtained using the
differential cross-section  (Eq.~\eqref{dsigma}) derived 
in Subsection~\ref{shape}, 
where the spectrum is given in the form of three dimensional integral,
that is easy to evaluate 
numerically with the 
Cuhre or Monte~Carlo algorithms. Additionally, we  express the 
moments of the spectrum as a combination of Fermi-Dirac integrals 
(Subsection~\ref{momenty}), that are used
to obtain convenient analytical
approximation of the spectrum in Subsect.~\ref{fit}.

Our long-term goal is to explore the entire neutrino spectrum produced
by massive stars at late nuclear burning stages. Unlike the Sun, these stars
 after carbon ignition
emit both neutrinos from weak nuclear reactions and thermal neutrinos.
Moreover, photon luminosity is negligible compared to neutrino luminosity and 
therefore these objects
are referred to as neutrino-cooled stars \cite{Arnett}. Neutrinos emitted
as neutrino-antineutrino pairs dominate up to the silicon burning 
\cite{MassiveStars}.
The electron antineutrinos are much easier to detect than neutrinos \cite{OMK, 
Epiphany, Gadzooks, 
LENA}. We consider them first. 

Production of 
pair-annihilation neutrinos is the dominant thermal process with relatively 
high
average neutrino energy of $\langle \mathcal{E}_{\nu} \rangle \sim$1.5~MeV. 
However, the information
on neutrino emission is not complete without full knowledge of all neutrino
 processes.
This is important because other neutrino processes can also produce low energy 
neutrinos
which are difficult to detect. These neutrinos somehow "steal" detectable 
energy from 
the stellar core reducing chances for e.g. supernova prediction.

The results presented here are general in nature, and are valid for pair-annihilation
neutrinos from $e^{+} e^{-}$ plasma in the full range of temperature and 
chemical potential.

\section{Pair annihilation in the Standard Model}

\begin{figure}
\begin{center}
\includegraphics[scale=0.4]{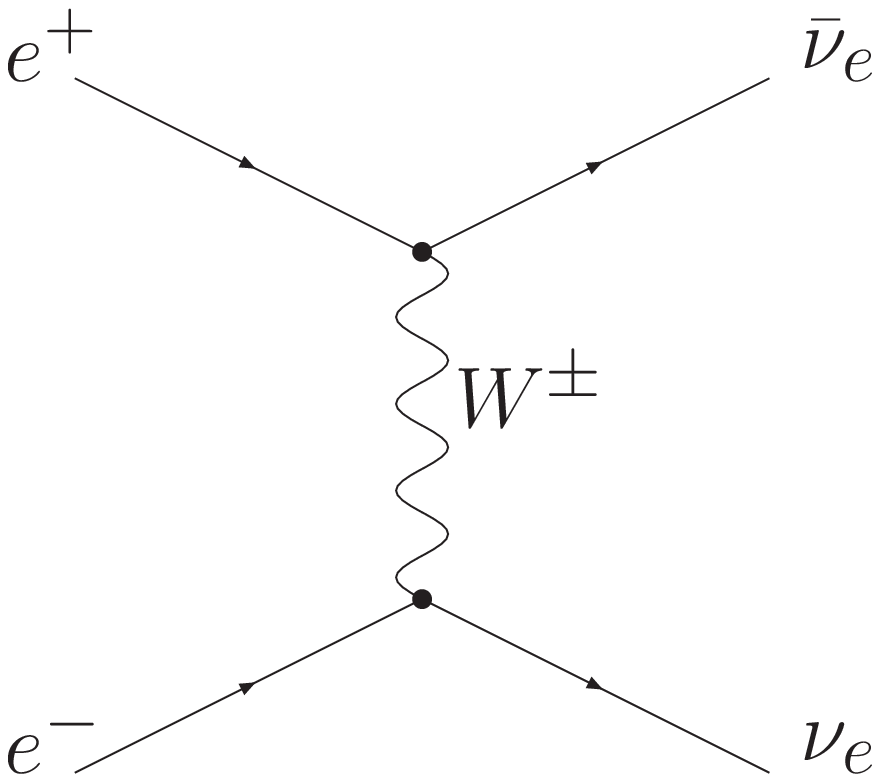}
$\quad$
\includegraphics[scale=0.4]{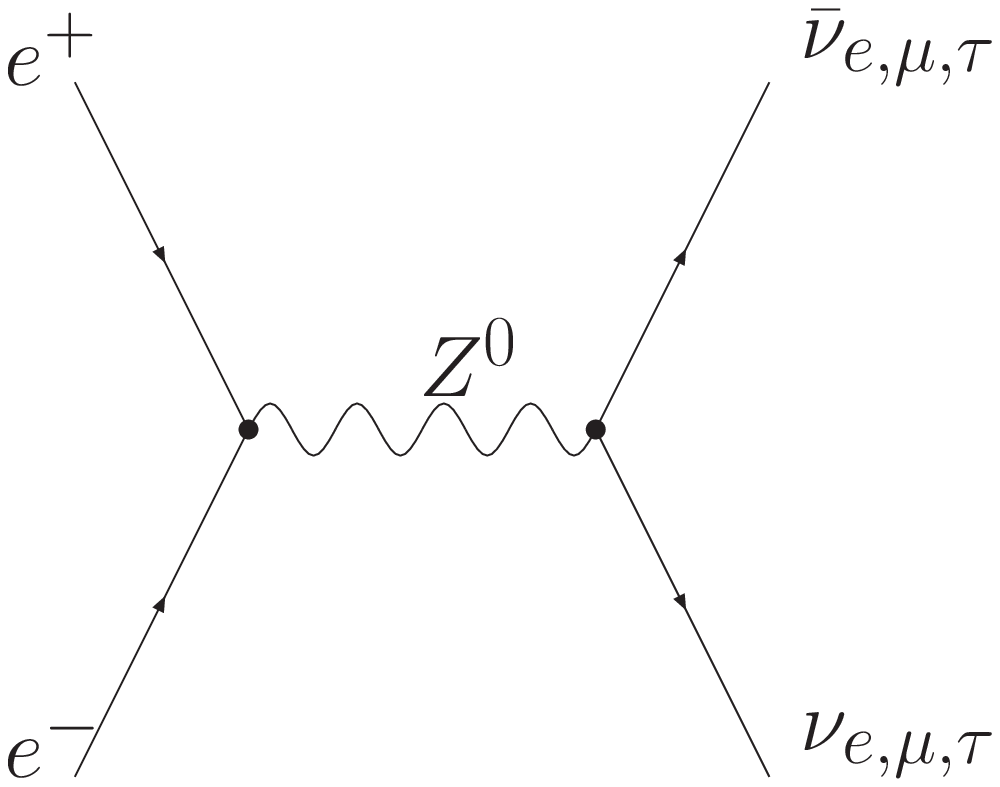}
\end{center}
\caption{Feynmann diagrams leading to the first-order amplitude for $e^+ e^-$ annihilation
into neutrinos. Electron neutrinos are produced via both diagrams, while
$\mu$ and $\tau$ neutrinos only by ``$Z^0$ decay''.}
\label{feynm}
\end{figure}

According to the Standard Model of electroweak interactions,
the electron-positron pair may annihilate not only into photons
but also into neutrino-antineutrino pair:
\begin{equation}
\label{eplus_eminus}
e^{+} + e^{-} \longrightarrow \nu_x + \bar{\nu}_x.
\end{equation}
In the first order calculations, sufficient in the considered energy range 
of several MeV 
($E\ll$~M$_{W^{\pm}, Z^0}$ where M$_{W^{\pm}, Z^0}$ is 
the intermediate boson mass of $\sim$100~GeV), two
Feynmann diagrams (Fig.~\ref{feynm}) contribute to the annihilation amplitude.

The contribution from the left diagram in Fig.~\ref{feynm}, i.e. $W^{\pm}$
boson exchange (charged current) is given by \cite{Dicus}:
\begin{equation}
\label{W_exchange}
-\frac{i G_F}{\sqrt{2}} \bar{u}_{\nu} \gamma^{\alpha} (1-\gamma_5) u_e
\;
\bar{v}_e \gamma_{\alpha} (1-\gamma_5) v_{\nu} .
\end{equation}
This part of the process produces only electron neutrinos. Using the Fierz 
transformation:
\begin{equation}
\bar{a} [\gamma^{\mu} (1-\gamma_5) ] b \;
\bar{c} [\gamma_{\mu} (1-\gamma_5) ] d
=
-\bar{a} [\gamma^{\mu} (1-\gamma_5) ] d \;
\bar{c} [\gamma_{\mu} (1-\gamma_5) ] b
\label{Fierz}
\end{equation}
we may write \eqref{W_exchange} as\footnote{We are dealing here with the field operators,
not numerical spinors, so the overall minus sign in Eq. (\ref{Fierz}) disappears \cite{Fierz_identities}.} :
\begin{equation}
\label{W_exchange_Fierz}
-\frac{i G_F}{\sqrt{2}} \bar{u}_{\nu} \gamma^{\alpha} (1-\gamma_5) v_{\nu}
\;
\bar{v}_e \gamma_{\alpha} (1-\gamma_5) u_e.
\end{equation}

The $Z^{0}$ boson exchange (neutral current) gives:
\begin{equation}
\label{Z_exchange}
-\frac{i G_F}{\sqrt{2}}
\bar{u}_{\nu} \gamma^{\alpha} (1-\gamma_5) v_{\nu}
\;
\bar{v}_e \gamma_{\alpha} ( g_V -  g_A\;\gamma_5) u_e \quad
\end{equation}
where:
$$
g_V= -\frac{1}{2} + 2 \sin^2{\theta_W}, \qquad g_A=-\frac{1}{2}
$$

By adding \eqref{W_exchange_Fierz} to  \eqref{Z_exchange} we obtain the total
 annihilation amplitude:
\begin{equation}
\label{annihilation_amplitude}
\mathcal{M}= -\frac{i G_F}{\sqrt{2}}
\bar{u}_{\nu} \gamma^{\alpha} (1-\gamma_5) v_{\nu}
\;
\bar{v}_e \gamma_{\alpha} (C_V^f - C_A^f \;\gamma_5) u_e.
\end{equation}
Here $G_F = 1.16637(1)\times 10^{-5} \, \mbox{GeV}^{-2}$ is the Fermi constant of 
weak interactions \cite{PDG}.
The parameters $C_V^f$ and  $C_A^f$ ($f=e, \mu ,\tau$ -- neutrino flavor) in the
 Standard Model
of the electroweak interactions are:\\
for electron neutrinos:
\begin{equation}
\label{CV_e}
C_V^e = \frac{1}{2} + 2 \sin^2{\theta_W}, \qquad C_A^e = \frac{1}{2}
\end{equation}
for $\mu$ and $\tau$ neutrinos:
\begin{equation}
\label{CV_mu}
C_V^{\mu, \tau} = -\frac{1}{2} + 2 \sin^2{\theta_W}, \qquad C_A^{\mu, \tau} = -\frac{1}{2},
\end{equation}

where $\theta_W$, the Weinberg angle, is $\sin^2{\theta_W}=0.22280 \pm 0.00035$ \cite{PDG}.

As the pair-annihilation into electron neutrinos proceeds via both charged and neutral currents
we have:
\begin{equation}
C_V^e = C_V^{\mu, \tau} + 1, \qquad C_A^e = C_A^{\mu, \tau}+1
\end{equation}
Annihilation into $\mu$ and $\tau$ neutrinos proceeds via the neutral current only.

In general, the entire spin-averaged information\footnote{If
electrons are polarized then we have to calculate $M^2$ using the density matrix for polarized
fermions \cite{polarized_fermions}.} on $e^+ e^{-}$ annihilation
 process \eqref{eplus_eminus}
is included in the spin-averaged ($\sigma_i$ - spin states) squared 
annihilation matrix $M^2$:
\begin{equation}
M^2 \equiv \overline{|\mathcal{M}|^{2}} = \frac{1}{2}\frac{1}{2} \sum_{\sigma_{e^{+}}, \sigma_{e^{-}}} \sum_{\sigma_{\nu}, \sigma_{\bar{\nu}}} |\mathcal{M}|^2.
\end{equation}

The amplitude $M^2$ can be calculated using e.g. the Casimir trick:
\begin{multline}
\label{M2}
M^2 = 8 {G_F}^2 \Big\{
\left(   C_A^f - C_V^f  \right)^2   P_1 \cdot Q_1\, P_2  \cdot Q_2 \\
+\left( C_A^f + C_V^f \right)^2     P_2 \cdot Q_1\, P_1  \cdot Q_2 \\
+ {m_e}^2 \left( {C_V^f}^2 - {C_A^f}^2 \right)\, Q_1 \cdot Q_2 \Big\} 	
\end{multline}
where $P_1 = (E_1, \mathbf{p}_1)$, $P_2 = (E_2, \mathbf{p}_2)$,
$Q_1 = (\mathcal{E}_1, \mathbf{q}_1)$  
$Q_2= (\mathcal{E}_2, \mathbf{q}_2)$ is, respectively, the four-momentum of the electron, positron,
neutrino and antineutrino.

According to the definition of the cross-section $d\sigma$, the number of
 collisions $dN$ occurring in
volume $dV$ in time $dt$ is \cite{Landau}:
\begin{equation}
\frac{dN}{dV dt} = d\sigma v \; d n_1\; d n_2 ,
\label{dn}
\end{equation}
where $d n_{1,2}$ are particle densities. 
In the case of two incoming particles, the differential cross-section can be 
computed from the Fermi 
Golden Rule formula
and is given by the expression \cite{Griffiths}:
\begin{equation}
d\sigma v = \frac{1}{2 E_1} \frac{1}{2 E_2}
\frac{1}{(2\pi)^2} \;
\delta^4(P_1 + P_2 - Q_1 - Q_2)\;
\frac{d^3 \mathbf{q}_1}{2 \mathcal{E}_1}
\frac{d^3 \mathbf{q}_2}{2 \mathcal{E}_2}\; M^2.
\label{do}
\end{equation}

Hence, if $M^2$ is known, the total neutrino emissivity from ${e^+} {e^-}$
 plasma can be calculated 
by performing appropriate integrations \cite{Hansel}:
\begin{equation}
\label{golden_fermi}
\frac{4}{(2\pi)^8}
\int
\frac{d^3 \mathbf{p}_1}{2 E_1}
\frac{d^3 \mathbf{p}_2}{2 E_2}
\frac{d^3 \mathbf{q}_1}{2 \mathcal{E}_1}
\frac{d^3 \mathbf{q}_2}{2 \mathcal{E}_2} 
\Lambda \;
f_1 f_2 \; \delta^4(P_1+\ldots) \;
M^2.
\end{equation}
Substituting into the integrand of \eqref{golden_fermi} the appropriate expression for 
the factor 
$\Lambda$ we obtain formulae for the number emissivity ($\Lambda=2$),
 the total emissivity 
(the energy carried by neutrinos and antineutrinos)
($\Lambda=E_1 + E_2 = \mathcal{E}_1 + \mathcal{E}_2$),  the antineutrino emissivity 
($\Lambda=\mathcal{E}_2$) etc. Functions $f_1, f_2$ are the Fermi-Dirac 
distributions
for electrons and positrons, respectively:
\begin{equation}
\label{F-D}
f_1 = \frac{1}{e^{(E_1 - \mu)/kT}+1}, \qquad
f_2 = \frac{1}{e^{(E_2 + \mu)/kT}+1},
\end{equation}
where $\mu$ is the electron chemical potential (including the rest mass) 
and neutrinos and antineutrinos
are assumed to escape freely from the plasma.

The total neutrino emissivity, as required e.g. for stellar evolution codes, 
including the emission of all three flavors ($e, \mu, \tau$) is the sum of three terms of 
the form of the
integral
\eqref{golden_fermi} with appropriate coefficients (\ref{CV_e}, \ref{CV_mu}) in \eqref{M2}.

The formal expression \eqref{golden_fermi} must be transformed into analytical 
or convenient numerical form  before actual application for e.g. stellar 
evolution 
codes or signal estimations in neutrino 
detectors. The integral \eqref{golden_fermi} above is simplified 
significantly if Lenard's formula \cite{Lenard, Lenard2}:
\begin{widetext}
\begin{multline}
\label{lenard}
\int \frac{d^3 \mathbf{q}_1}{2\,\mathcal{E}_1} \frac{d^3 \mathbf{q}_2}{2\,\mathcal{E}_1}
\;
Q_1^{\alpha} Q_2^{\beta}
\;
\delta^4(P_1 +P_2 - Q_1 -Q_2)
= 
\frac{\pi}{24}
\left[
g^{\alpha \beta}(P_1 + P_2)^2
+2\; ( P_1^{\alpha} + P_2^{\alpha})
(P_1^{\beta} + P_2^{\beta})
\right] \Theta \left[ (P_1 + P_2)^2 \right]
\end{multline}

is applied to integrate over outcoming neutrino momenta giving the 
well-known cross 
section\footnote{We denote expression \eqref{ovDicus} as ${\sigma v}_D$ to avoid confusion with 
differential cross-sections (\ref{dsigma}).}
of \citet{Dicus} for annihilation of the $e^- e^+$ pair with four-momenta $P_1$ and $P_2$:

\begin{equation}
\label{ovDicus}
{\sigma v}_D = \frac{{G_F}^2}{12 \pi} \frac{ {m_e}^4 }{E_{1}  E_{2}}
\Biggl [  
\left({C_V^f}^2 + {C_A^f}^2  \right)   \left (  1+ 3\; \frac{P_1 \cdot P_2}{{m_e}^2} +2\;
\frac{(P_1 \cdot P_2)^2}{{m_e}^4}  \right )
+
\left({C_V^f}^2 - {C_A^f}^2  \right)  \left(  1+ 2\; \frac{P_1 \cdot P_2}{{m_e}^2}   \right)
\Biggr ]
\end{equation}
\end{widetext}

The formula \eqref{golden_fermi} can be now transformed using \eqref{ovDicus} into:
\begin{equation}
\label{total_emi}
\frac{4}{(2\pi)^6} \int d^3 \mathbf{p}_1 d^3 \mathbf{p}_2 \;  \Lambda\;  {\sigma v}_D \; f_1 f_2 .
\end{equation}
In contrast to a more general Eq.~\eqref{golden_fermi}, in \eqref{total_emi} $\Lambda$ must be 
a function of $e^-, e^+$ momenta $\mathbf{p}_1, \mathbf{p}_2$ only, and we are unable
to compute the antineutrino spectrum separately.
The expression \eqref{total_emi} above is actually a three-dimensional integral -- due to the
rotational symmetry only lengths and angle between $\mathbf{p}_1$ and $\mathbf{p}_2$ are independent
variables of integration. Therefore, \eqref{total_emi} may be integrated, 
leading to a combination of Fermi-Dirac integrals (cf. Sect.~\ref{avg}) which are  easily 
computed numerically. 

We can obtain the total neutrino luminosity in this way.
Unfortunately, if one attempts to consider the detection of these neutrinos
very important information about the neutrino energy is missing. Detection efficiency 
depends 
strongly on the neutrino energy -- higher energy neutrinos are much easier to 
detect -- so the 
spectrum parameters and the shape actually decide whether these neutrinos will be 
observed or not. 
Therefore we must go back to the general formula \eqref{golden_fermi} and try to 
attack it 
in an another way. 
We consider various possible approaches in the next subsections.

\section{Neutrino spectrum}

\subsection{Monte~Carlo simulations \label{monte}}

\begin{figure}
\includegraphics[angle=270,width=8.6cm]{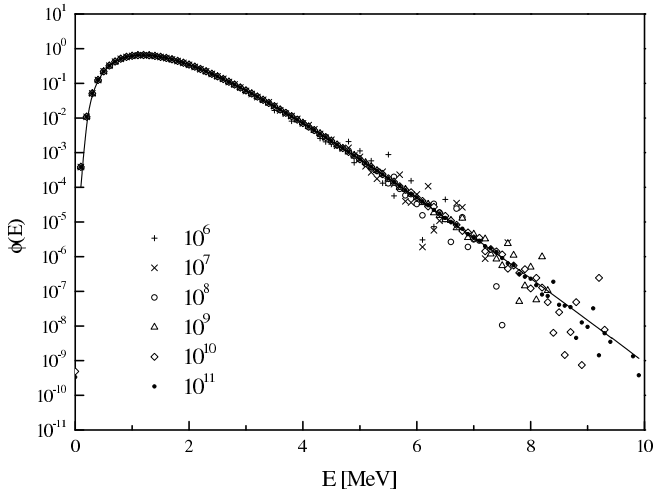}
\caption{\label{monte_vs_cuhre} Spectrum resulting from the Monte~Carlo simulation for $10^6, 10^7, 10^8, 10^9, 10^{10}$ and $10^{11}$ loop size (symbols) \textit{versus} spectrum from \eqref{cuhre_widmo} computed with aid of Cuhre algorithm (solid line, relative accuracy 0.05). 
We can notice decreasing 
errors of the Monte~Carlo simulation near the maximum and increasing number of events
in the high energy tail. Obviously, the Monte~Carlo simulation is unable to reach
 accuracy
of Cuhre methods for $E\gg$5~MeV.
This figure clearly justifies efforts to compute the spectrum using the 
appropriate cross-section
 \eqref{dsigma}, instead of using Monte~Carlo simulation based on the knowledge of the annihilation 
 matrix \eqref{M2}.
We would like to point out that three-dimensional integral \eqref{cuhre_widmo} may also
 be computed 
using advanced Monte~Carlo adaptive algorithms, giving similar spectra much faster, but they 
may fail in some cases, cf.~Fig.~\ref{cuhre_figure}. 
Spectrum computed for typical situation in stellar center during Si burning with
$T=0.319$~MeV and $\mu=0.85+m_e$~MeV,
}
\end{figure}

The Monte~Carlo simulation is a method often used to calculate annihilation 
spectra. We applied 
it in our previous work \cite{OMK}, with some minor errors in the algorithm. 
To improve the 
calculations we develop semi-analytical methods, to make cross-check of 
the obtained results. 
Poor accuracy for higher energies of previously obtained results 
did not allow us 
to compute reliably the spectrum for energies E$>$4~MeV.
Using better random number generator, namely Mersenne Twister routine \cite{mt}, 
and longer 
simulation runs, we were able to  extend the spectrum to 6-7~MeV at least,
but accuracy was still unacceptable, cf.~Fig.~\ref{monte_vs_cuhre}.

As we can see in Fig.~\ref{monte_vs_cuhre} the Monte~Carlo simulation reproduces the spectrum well.
Unfortunately, this method is very slowly convergent: time required to compute the grid of neutrino
spectra for e.g. the purpose of integrating them
over the entire volume of the hot stellar core is very long. Moreover, the accuracy of this method is 
limited
by resolution of the random number generator, and errors in the interesting range
above $\sim$5~MeV (Super-Kamiokande threshold) are large. Therefore development
of the analytical formulae is welcome. Nevertheless, Monte~Carlo simulations based only on the amplitude
$M^2$ \eqref{M2} may be used even in the case where analytical formula for the cross-section 
is not available.

\subsection{Emissivity and average neutrino energy \label{avg}}

Some quantities related to the neutrino emission process, such as reaction rates, the total emissivity 
and the average neutrino energy, can be easily obtained directly from 
the Dicus cross-section \eqref{ovDicus} as electron energy moments. These quantities are average values 
for $\nu-\bar{\nu}$ pair. However, in Sect.~\ref{shape} we show how to modify average values 
by small additive terms to obtain the relevant quantities for neutrinos and antineutrinos separately. 
This becomes possible only using a more general cross-section \eqref{dsigma}. 

Our intermediate goal is to compute the integral:
\begin{equation}
I_{n\;m} = \frac{4}{(2 \pi)^6}    \int d^3\mathbf{p}_1 d^3\mathbf{p}_2 \; {\sigma v}_D   \; {E_1}^n {E_2}^m\; f_{1} f_{2}
\end{equation}
i.e. the expression \eqref{total_emi} with $\Lambda = {E_1}^n {E_2}^m$
where the cross-section ${\sigma v}_D$ is given by the formula \eqref{ovDicus}, and $n, m$ are arbitrary 
real numbers.

The integral over cosine of the angle between $\mathbf{p}_1$ and  $\mathbf{p}_2$  ($\cos{\theta}$) 
can be easily done, while integrals
over $dE_1 dE_2$ can be separated and expressed by the products of  Fermi integrals.
This is a result of the fortunate coincidence: the expressions $\sqrt{E^2-{m_e}^2}$ appear
(i) from phase-space factors $d^3 \mathbf{p}$ and (ii) from four-momentum products,
but the latter ones are always squared because the integration of $\cos^n \theta$ with an odd power 
gives zero.

Let us define functions

\begin{multline}
M_{\mp}^{nm} =
  (7 {C_V}^2 - 2 {C_A}^2)\; G_{n/2-1/2}^{\mp} G_{m/2-1/2}^{\pm}\\
+9 {C_V}^2 \; G_{n/2}^{\mp} G_{m/2}^{\pm}+
({C_V}^2 + {C_A}^2)\; \Bigl (
4\, G_{n/2+1/2}^{\mp} G_{m/2+1/2}^{\pm}\\ - G_{n/2-1/2}^{\mp} G_{m/2+1/2}^{\pm} -G_{n/2+1/2}^{\mp} G_{m/2-1/2}^{\pm}
\Bigr )
\end{multline}

where the Fermi integrals $G_n^{\pm}$ are:
\begin{equation}
\label{FD}
G_n^{\pm}(\alpha, \beta) =
\frac{1}{\alpha^{3+2 n}} \int_{\alpha}^{\infty} \frac{x^{2 n+1 } \sqrt{x^2-\alpha^2}}{1+\exp(x\pm \beta)} \; dx
\end{equation}
$\alpha = m_e/kT$, $\beta = \mu_e/kT$, $x=E/kT$.

The functions $M_{\mp}^{nm}$ obey the relation:
\begin{equation}
M_{-}^{nm} = M_{+}^{mn}
\end{equation}

For example, n-th moment of the electron energy is proportional to
$M_{-}^{n 0}$ and n-th moment of the positron energy is proportional to $M_{+}^{n 0} = M_{-}^{0 n}$

Electron (positron) energy moments can thus be calculated as:
\begin{equation}
\label{moments}
I_{nm} =  \frac{ {G_F}^2 {m_e}^{8+n+m} }{18 \pi^5}  M_{-}^{n m}
\end{equation}

We can now write elegant expressions for the emissivities, which are equivalent to
well-known formulae \cite{Itoh2}, \cite{Dicus}, \cite{bisnovaty}.
The total emissivity (the neutrino energy produced per unit volume and unit time) is equal to:
\begin{equation}
\label{Q}
\frac{d E}{dV dt} \equiv Q = \frac{ {G_F}^2 {m_e}^{9} }{18 \pi^5}  \left ( M_{-}^{1 0}+M_{+}^{1 0} \right )
\end{equation}
while the number emissivity (particle production rate) is:
\begin{equation}
\label{F}
\frac{d N}{dV dt} \equiv F  = 2 R =  \frac{ {G_F}^2 {m_e}^{8} }{18 \pi^5} (M_{-}^{0 0} + M_{+}^{0 0})
\end{equation}
The reaction rate $R$ is half of the particle  emissivity as two neutrinos
are produced in a single event \eqref{eplus_eminus}. 
In eq.~\eqref{Q} energies of two neutrinos ($\nu$ and $\bar{\nu}$) are added together.

We can also compute the average neutrino energy from the pair-annihilation process:
\begin{equation}
\label{E_average}
\langle \mathcal{E} \rangle =  \frac{Q}{F} = \frac{Q}{2 R}  \quad .
\end{equation}

Average neutrino energy as a fraction of the electron rest-energy $m_e c^2$ is:
\begin{equation}
\label{total_avg}
\langle \mathcal{E} \rangle = \frac{ M_{-}^{1 0}+M_{+}^{1 0} }{M_{-}^{0 0}+M_{+}^{0 0}} \;\; [ m_e c^2 ].
\end{equation}
Unfortunately, no more informations of interest on the neutrino spectrum can be extracted from 
Lenard-formula based calculations. We wish to point out that the equation \eqref{total_avg} gives the 
average $\nu_x$-$\bar{\nu}_x$ energy 
$\langle \mathcal{E} \rangle = (\langle \mathcal{E}_1 \rangle + \langle \mathcal{E}_2 \rangle)/2$. 
However, both mean energies $\langle \mathcal{E}_1 \rangle$ and $\langle \mathcal{E}_2 \rangle$
are not identical, but differ slightly. They can be both computed from eq.~\eqref{cuhre_moments}.

Knowledge of the mean neutrino energy from the formula \eqref{total_avg}  allows one to quickly estimate 
chances for neutrino 
detection from a given reaction. At present, neutrinos
of a few keV energy are impossible to detect, 1~MeV neutrinos are difficult, while for 10~MeV neutrinos
we have mature detection techniques with megaton targets soon available.

Using analytical approximations for Fermi-Dirac integrals \cite{BlinnikovRudzskij, BlinnikovRudzskij2}, 
we derive useful analytical formulae for average neutrino energy. Particularly simple expressions exist
in the following cases:
\begin{itemize}
\item{Relativistic and non-degenerate plasma 
$kT > 2\,m_e$, $kT>\mu$:
\begin{equation}
\label{R_ND}
\langle \mathcal{E} \rangle = \frac{2700}{7} \frac{\zeta (5)}{\pi^4}\; kT = 4.106\; kT \sim 4\; kT
\end{equation}
}
\item{Non-relativistic and non-degenerate plasma \mbox{$kT \ll m_e$}, $kT>\mu$:
\begin{equation}
\label{NR_ND}
\langle \mathcal{E} \rangle = m_e + \frac{3}{2}\, kT
\end{equation}
}
\item{Degenerate case $\mu \gg kT$:
\begin{equation}
\label{D}
\langle \mathcal{E} \rangle = \frac{2}{5}\; \mu + 2\; kT + \frac{m_e}{2}
\end{equation}
}
\end{itemize}

Unfortunately, in the central region of a massive star  $m_e \simeq kT \simeq \mu$
and none of the cases above holds. Eq.~\eqref{total_avg} must be evaluated
numerically.

\subsection{Spectrum shape \label{shape}}

To find the flux of more-than-average energy neutrinos we have to compute the neutrino spectrum. To do this
we go back to the formula \eqref{golden_fermi}, and integrate it step-by-step.
Without the Lenard's formula \eqref{lenard} these calculations are ``tedious algebra'' 
\cite{YuehBuhler}, \cite{HannestadMadsen} and lead to very complicated expressions.

We have computed the essential differential cross section for $e^+ e^-$ annihilation into neutrino 
(${d \sigma v}/{d \mathcal{E}_{1}}$) or antineutrino (${d \sigma v}/{d \mathcal{E}_{2}}$) of energy 
$\mathcal{E}_{1,2}$ measured in the plasma rest frame:

\begin{widetext}
\begin{equation}
\label{dsigma}
\frac{d \sigma v}{d \mathcal{E}_{1,2}} = \frac{{G_F}^2}{8 \pi \;E_{1} E_2} \; 
\left [ 
(C_V+C_A)^2 \; H_{1,2} + (C_V-C_A)^2 \; H_{2,1} + 2\, {m_e}^2\; ({C_V}^2+{C_A}^2) H_3
\right ]
\varTheta
\end{equation}
\end{widetext}

where $\varTheta$ is 
(
$\mathcal{E}_{i=1,2}$, $\mathcal{E}_1$ - $\nu$ energy, $\mathcal{E}_2$ - $\bar{\nu}$  energy
)
\begin{equation}
\label{UnitStep}
\varTheta =
\begin{cases}
0 &  \text{for} \qquad \mathcal{E}_i < \mathcal{E}_{-}\\
1 &  \text{for} \qquad \mathcal{E}_{-} <  \mathcal{E}_i < \mathcal{E}_{+}\\
0 &  \text{for} \qquad \mathcal{E}_i > \mathcal{E}_{+}
\end{cases}
\end{equation}
and:
\begin{equation}
\mathcal{E}_{\pm} = \frac{1}{2} \left( E_{1} + E_2 \right) \pm \frac{1}{2} \left| \mathbf{p}_1 + \mathbf{p}_2 \right|.
\end{equation}
These values ($\mathcal{E}_{-}$,$\mathcal{E}_{+}$) restrict the neutrino energy $\mathcal{E}_i$ to 
kinematically allowed (in a single $e^+ e^-$ annihilation event) values.

The neutrino cross-section is different from the antineutrino one, and the
formula for neutrinos can be obtained from expression for antineutrinos 
by exchange of $H_1$ and $H_2$ in \eqref{dsigma}, as it is indicated by subscripts.

Full expressions for $H_1$, $H_2$ and $H_3$ in (\ref{dsigma}) are complicated:
\begin{equation}
H_k = \sum_{j=1}^5  h_k^j \left| \mathbf{p}_1 + \mathbf{p}_2 \right|^{-j}.
\end{equation}

With the index $i=1,2$ corresponding to neutrino and antineutrino, respectively, and  
$$
p_1 \, ( p_1 + p_2 \, \cos{\theta} ) = \delta, \qquad 
\mathcal{E}_i \; (E_1 + E_2)  - P_1 \cdot P_2 - {m_e}^2 = \Delta
$$
non-zero $h_i^j$ are:
\begin{subequations}
\begin{eqnarray*} 
h_1^1&=& 2 \,(\Delta - \mathcal{E}_i E_2)^2 \\
h_1^3 &=& - 4 \delta (\Delta - \mathcal{E}_i E_2) \Delta
- p_1^2 \Delta^2 
+ {\mathcal{E}_i}^2 {p_1}^2 {p_2}^2 \sin^2{\theta}\\
h_1^5 &=& 3\, \delta \, \Delta^2\\
h_2^1 &=& 2 {E_1}^2 {\mathcal{E}_i}^2\\
h_2^3 &=& - 4 \mathcal{E}_i E_1 \delta \Delta
- {p_1}^2 \;\Delta^2 +  {\mathcal{E}_i}^2 {p_1}^2 {p_2}^2 \sin^2{\theta} \\
h_2^5 &=&  3\, \delta \, \Delta^2\\
h_3^1 &=& \Delta - \mathcal{E}_i (E_1 + E_2)
\end{eqnarray*}
\end{subequations}

where $p_i\equiv|\mathbf{p}_i|$ denotes the length of 3-momentum and 
\mbox{$\theta = \angle(\mathbf{p}_1,\mathbf{p}_2)$} is the angle between incident electron and positron.

The well-known formula for Dicus cross-section \eqref{ovDicus} is reproduced 
by computing either of the following integrals:
\begin{equation}
\label{ov_int}
{\sigma v}_D = \int_{\mathcal{E}_{-}}^{\mathcal{E}_{+}} \quad  \frac{d \sigma v} {d\mathcal{E}_{1,2} }  \;
\; d\,\mathcal{E}_{1,2}.
\end{equation}

The neutrino and antineutrino spectrum may be computed from the following formula:
\begin{equation}
\label{cuhre_widmo}
\lambda(\mathcal{E}_{1,2}) = \mathcal{N} \; \int d^3\mathbf{p}_1 d^3\mathbf{p}_2 \; \frac{d \sigma v}{d \mathcal{E}_{1,2}}  f_{1} f_{2} 
\end{equation}
where the normalization constant is related to the reaction rate,
\begin{equation}
\mathcal{N} = R^{-1}, 
\end{equation}
and can be computed from the formula \eqref{F}. Reaction rate
can be computed from the cross-section \eqref{dsigma} as well, but due to \eqref{ov_int}
these expressions are equal. This is also obvious from physical point of view:
the number of reactions is equal to number of  neutrinos (or antineutrinos) emitted.

The integral \eqref{cuhre_widmo} above, because of the presence of the unit step function in 
the cross-section \eqref{dsigma}, must be evaluated numerically\footnote{Our code {\tt PSNS \label{psns}} 
\cite{PSNS} 
will be available at {\tt http://th-www.if.uj.edu.pl/psns}.}.

The most reliable method for multidimensional integration is the Cuhre algorithm, but
it is also possible to calculate integrals \eqref{cuhre_widmo} using
Monte~Carlo algorithms. In actual calculations we  used the 
Cuba library \cite{libcuba}. To get the complete neutrino spectrum,
one has to compute numerically three-dimensional integral at every point.

\begin{figure}
\includegraphics[angle=270,width=8.6cm]{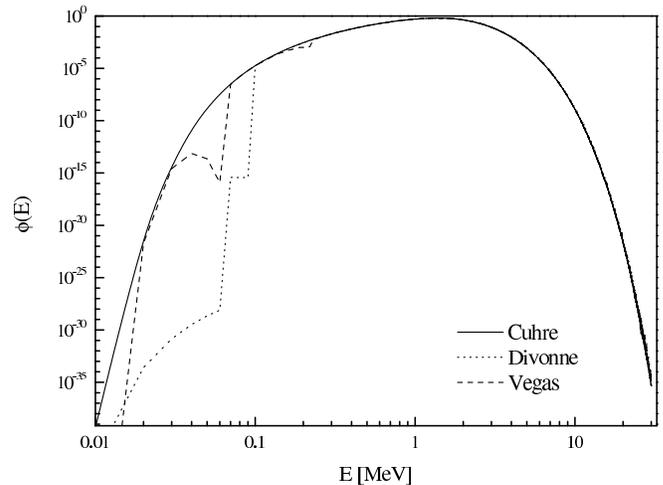}
\caption{
\label{cuhre_figure}
Spectrum of the $\mu$ and $\tau$ antineutrinos, 
computed from \eqref{cuhre_widmo}
with the use of the Cuhre algorithm implemented in Cuba library \cite{libcuba}
used in our PSNS code \cite{PSNS}. Guaranteed relative 
accuracy is everywhere better than 3\%.
This figure explains why Cuhre deterministic algorithm (solid line), in spite of slow convergence, is 
recommended to compute neutrino spectrum. Failure of Monte~Carlo algorithms (dashed and dotted line)
is apparent. However, these failures do not influence energy moments (mean energy and
dispersion of the spectrum) significantly leading to errors smaller than those Monte~Carlo algorithm 
produces itself.
Detailed knowledge of the spectrum below 0.1~MeV seems also unimportant from experimental point of view. Temperature and chemical potential values as in Fig.~\ref{monte_vs_cuhre}. 
}
\end{figure}

\begin{figure}
\includegraphics[angle=270,width=8.6cm]{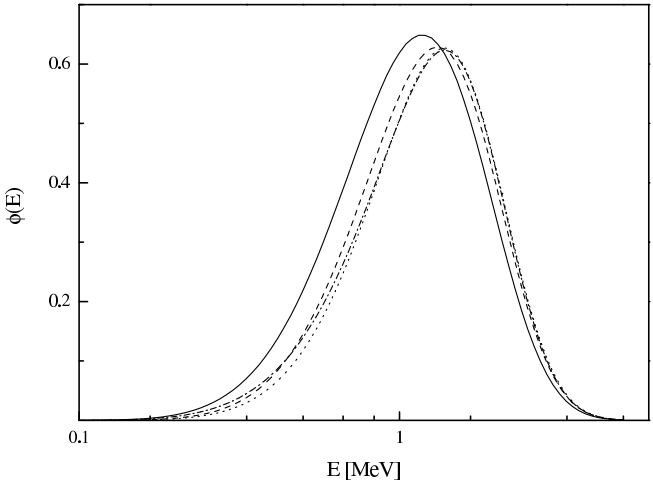}
\caption{\label{4_flavours} On semilog plot differences between neutrino spectra of electron neutrinos (dashed),
electron antineutrinos (solid), $\mu$ ($\tau$) neutrinos (dotted) and $\mu$ ($\tau$) antineutrinos (dash-dotted)
are clearly visible.  All four spectra are normalized to 1. $T$ and $\mu$ the same as in Fig.~\ref{monte_vs_cuhre}.
}
\end{figure}

Sample results produced by our code \cite{PSNS} are given in Fig.~\ref{cuhre_figure}
and in Fig.~\ref{4_flavours}. Noteworthy, Monte~Carlo integration apparently fails
to compute spectrum for neutrino energy $\mathcal{E}<0.1$~MeV. However, taking
into account predicted purpose of calculations, these errors are
insignificant. Typically, spectrum will be used to estimate
signal in neutrino detectors. Energy threshold for neutrino
detection is usually much higher than 0.1~MeV, and this part of the spectrum is not
detected at all. Therefore we may take advantage of the Monte~Carlo integration
performance, and compute spectrum much more than ten times faster. For 
theoretical considerations, we however recommend use of much reliable Cuhre
algorithm.

In Fig.~\ref{4_flavours} we can see clearly differences between
$\nu_e$, $\bar{\nu}_e$, $\nu_{\mu, \tau}$ and $\bar{\nu}_{\mu, \tau}$ neutrino
spectra. Particularly, electron anti-neutrino spectrum,
main goal of e.~g. \textit{GADZOOKS!} detector \cite{Gadzooks} has
lower mean energy and tail than other neutrinos.

Quality of the results computed using PSNS \cite{PSNS} may be judged
using data from Table~\ref{num_vs_anal}, where obtained numerically spectrum
was integrated to get average neutrino energy (columns 4-7). Effectively this
means that 4-dimensional integrals have been computed. Mean energies
were computed again using results from Sect.~\ref{momenty}, as
combination of Fermi-Dirac integrals \eqref{FD} with accuracy
of at least $10^{-7}$ (columns 8-11, only 4 digits are shown). No significant
discrepancies were found. Moreover, for $\mu \ll kT$ or $\mu \gg kT$,
asymptotic expansions \eqref{R_ND} and \eqref{D}, respectively (columns 2-3),
may be used as a very good estimate.

\begin{table*}
\caption{\label{num_vs_anal}
Comparison of the average neutrino energy $\langle \mathcal{E} \rangle$
computed from numerically calculated (using \eqref{cuhre_widmo} implemented in \cite{PSNS}) spectrum
and formulae (\ref{R_ND}--\ref{D}). Chemical potential was chosen to be equal to $0.85+m_e$~MeV.
Columns 2 and 3 show results
obtained from asymptotic expansions \eqref{D} and \eqref{R_ND}. Up to the first order
no differences between electron and mu/tau flavor exist. Columns 4-7
show mean neutrino energy as computed using spectrum from our PSNS \cite{PSNS} code
employing CUHRE deterministic multidimensional integration algorithm from Cuba \cite{libcuba} library.
Columns 8-11 show results computed from our formulae \eqref{total_avg} (mean neutrino-antineutrino pair energy)
corrected using \eqref{deltaQ} (mean neutrino energy is, of course $Q/R$ where reaction rate is given
by \eqref{F}) to get separately neutrino and antineutrino energies. Results are in perfect
agreement (compare columns 4-7 and 8-11) even up to 4 digits. As Fermi-Dirac integrals \eqref{FD}
(used in calculations presented in columns 8-11)
are easily computed with accuracy of at least several digits \cite{FDnum1, FDnum2}, we may conclude,
that our code \cite{PSNS} performs very well, and produce reliable results
with accuracy possibly much better than 5\% as estimated by Cuba library algorithms.
Our analytical expansions (\ref{D}, \ref{R_ND})	are very useful (cf. columns 2-3 \textit{versus}
column 12 or 13) in the appropriate (degenerate or non-degenerate relativistic) conditions,
as long as one do not need to know about differences between neutrino flavor, i.e neutrino-antineutrino
asymmetry and $e$-$\mu,\tau$ asymmetry.
}
\begin{tabular}{c||cc|cccc|cccc|cc}
\parbox{2cm}{$kT$  [MeV] \\($\mu=1.361$) }
& \parbox{2cm}{$\frac{2}{5}\mu + 2kT$\\$+m_e/2$} 
& $4.106\; kT$
& $\langle \mathcal{E}_{\nu_e} \rangle$
& $\langle \mathcal{E}_{\bar{\nu}_e} \rangle$ 
& $\langle \mathcal{E}_{\nu_{\mu,\tau}} \rangle$ 
& $\langle \mathcal{E}_{\bar{\nu}_{\mu,\tau}} \rangle$
& $\langle \mathcal{E}_{\nu_e} \rangle$
& $\langle \mathcal{E}_{\bar{\nu}_e} \rangle$ 
& $\langle \mathcal{E}_{\nu_{\mu,\tau}} \rangle$ 
& $\langle \mathcal{E}_{\bar{\nu}_{\mu,\tau}} \rangle$
& $\langle \mathcal{E}_{\nu_e-\bar{\nu}_e} \rangle$ 
& $\langle \mathcal{E}_{{\nu}_{\mu,\tau}-\bar{\nu}_{\mu,\tau}} \rangle$ \\
\hline
\hline
0.01 & 0.812 & 0.04 
& 0.703  & 0.913  & 0.753  & 0.911
& 0.7037 & 0.9171 & 0.7532 & 0.9168 & 0.810 & 0.835\\
\hline
0.05 & 0.900 & 0.205 
& 0.759  & 0.955  & 0.822  & 0.946
& 0.7584 & 0.9544 & 0.8211 & 0.9469 & 0.856 & 0.884\\
\hline
0.1  & 1.000 & 0.411
& 0.852  & 1.029 & 0.930  & 1.023
& 0.8510 & 1.028 & 0.9291 & 1.024   & 0.940 & 0.977\\
\hline
0.5  & 1.800 & 2.053
& 2.149 & 2.267 & 2.2495 & 2.2848
& 2.1494 & 2.2672 & 2.2496 & 2.2854 & 2.208 & 2.268\\
\hline
1.0  & 2.800 & 4.106
& 4.1260 & 4.2329 & 4.2043 & 4.2333
& 4.1277 & 4.2349 & 4.2050 & 4.2344 & 4.181 & 4.220\\
\hline
2.0 & 4.800 & 8.212 
& 8.1912 & 8.2952 & 8.2507 & 8.2781
& 8.1968 & 8.3004 & 8.2557 & 8.2833 & 8.249 & 8.270\\
\hline
5.0 & 10.80 & 20.53
& 20.476 & 20.578 & 20.525 & 20.551
& 20.495 & 20.597 & 20.540 & 20.568 & 20.546 & 20.554\\
\hline
10.0 & 10.80 & 41.06
& 40.97 & 41.08 & 41.02  & 41.05
& 41.019 & 41.121  & 41.053 & 41.081 & 41.07 & 41.07\\
\hline
100.0 & 100.8 & 410.6 
& 410.1 & 410.2 & 410.1 & 410.1 
& 410.56 & 410.66 & 410.60 & 410.62  & 410.61 & 410.61\\
\hline
\end{tabular}
\end{table*}

\subsection{Neutrino energy moments \label{momenty}}

Fortunately, we were lucky to express moments of the neutrino spectrum
by the Fermi-Dirac integrals \eqref{FD}. With the moments given, one is able to approximate
the spectrum with the aid of an appropriate analytical formula.

Neutrino and antineutrino energy moments are computed as integrals
\begin{equation}
\label{cuhre_moments}
J_n^{1,2} = 
\int_{\mathcal{E}_{-}}^{\mathcal{E}_{+}} d\mathcal{E}_{1,2} \int d^3\mathbf{p}_1 d^3\mathbf{p}_2 \; \frac{d \sigma v}{d \mathcal{E}_{1,2}}  f_{1} f_{2} \;
{\mathcal{E}_{1,2}}^n .
\end{equation}

Unexpectedly, integration over the neutrino energy $d \mathcal{E}_1$
and the angle between $\mathbf{p}_1$ and $\mathbf{p}_2$ in \eqref{cuhre_moments} can be done 
analytically for any integer value\footnote{At least up to $n=4$
where we stopped calculations.} of $n$. However, we are unable to find
general expression valid for any $n$ similar to Eq.~\eqref{moments}.

For $n=0$, due to \eqref{ov_int}, Eq.~\eqref{cuhre_moments} reduces to Eq.~\eqref{moments}
with $n=m=0$.
Physically, it expresses the fact, that the reaction rate is equal to the number of emitted neutrinos or 
antineutrinos per unit volume and per unit time.

For $n=1$, Eq.~\eqref{cuhre_moments} is equal to the neutrino (antineutrino)
emissivity. For convenience, we express it in the form:
\begin{equation}
J_1^{1,2} \equiv Q_{1,2} = \frac{Q}{2} \mp \Delta Q,
\end{equation}

where $Q$ is the total emissivity \eqref{Q} and $\Delta Q$ is:
\begin{multline} 
\label{deltaQ}
\Delta Q = \frac{{G_F}^2}{36 \pi}\; C_V C_A \; 
\Bigr [ 
4 \left ( G_1^{-}  G_{1/2}^{+} -  G_1^{+}  G_{1/2}^{-} \right) \\+ 
4 \left ( G_0^{-}  G_{-1/2}^{+} -  G_0^{+}  G_{-1/2}^{-} \right) -
\left ( G_1^{-}  G_{-1/2}^{+} -  G_1^{+}  G_{-1/2}^{-} \right)\\ -
7 \left ( G_0^{-}  G_{1/2}^{+} -  G_0^{+}  G_{1/2}^{-} \right) 
\Bigl ]
\end{multline} 

From \eqref{deltaQ} we can see explicitly when neutrino and antineutrino emissivities 
(as well as the spectra) are not identical. This happens for
\begin{enumerate}
\item{Parity violating interaction  i.e. both $C_V$ and $C_A$ in \eqref{annihilation_amplitude}
 must be non-zero}
\item{$G_n^{+} \neq G_n^{-}$ i.e. distributions of electrons and positrons must be different. As the only 
difference between $G_n^{+}$ and $G_n^{-}$ comes from chemical potential (cf.~Eq.~\eqref{FD}),  $\mu_e$ must not be negligible.}
\end{enumerate}

However, relative difference between neutrino and antineutrino emissivity usually is very small.
For example, in  the non-degenerate and relativistic case the total emissivity is:
\begin{equation}
Q = \frac{7\;  \zeta (5) }{12\; \pi}\; {G_F}^2\; \left ({C_V}^2 + {C_A}^2 \right) (kT)^9
\end{equation}
where $7\zeta(5)/(12 \pi)=0.1925$, while the difference between $\nu$ and $\bar{\nu}$ luminosity is only:
\begin{equation}
2\,\Delta Q = \left( \frac{49\, \pi^3}{8100} -  \frac{45\, \zeta(3) \, \zeta(5)  }{\pi^5} \right)\; 
{G_F}^2 \; C_V\, C_A\; \mu \; (kT)^8
\end{equation}
with the numerical coefficient in the parentheses equal to $0.0043$.

We also present the second moment of the neutrino spectrum. Again, the formula is split into
the average value $\langle J_2 \rangle$ and a small additive term $\Delta J_2$:
\begin{equation}
J_2^{1,2} = \langle J_2 \rangle \pm \Delta J_2.
\end{equation}

The average second moment is equal to:
\begin{multline}
\label{J2}
\langle J_2 \rangle = \frac{ {G_F}^2 }{ 360 \pi } \;
\Biggl \{
({C_V}^2+{C_A}^2) \;
	\Bigl [
		28\, \left( G_{3/2}^{-} G_{1/2}^{+} + G_{3/2}^{+} G_{1/2}^{-} \right)
		\\+
		30\, G_{1}^{-} G_{1}^{+} 
		-
		7\,  \left( G_{3/2}^{-} G_{-1/2}^{+} + G_{3/2}^{+} G_{-1/2}^{-} \right)
		\\-
		8\,  \left( G_{1/2}^{-} G_{-1/2}^{+} + G_{1/2}^{+} G_{-1/2}^{-} \right)
	\Bigr ] 
	+\\
	{C_V}^2
	\Bigl [
		70\, \left( G_{1/2}^{-} G_{-1/2}^{+}  + G_{1/2}^{+} G_{-1/2}^{-}  \right)
		+
		60\, \left( G_{1}^{-} G_{0}^{+}  + G_{1}^{+} G_{0}^{-}  \right)
	\Bigr ]
	\\+
	30\, ({C_V}^2 - {C_A}^2) G_{0}^{+} G_{0}^{-} 
	+
	16\, (3 {C_V}^2 - 2 {C_A}^2) G_{1/2}^{-} G_{1/2}^{+} 
	\\-
	(34 {C_V}^2 - 6 {C_A}^2)   G_{-1/2}^{-} G_{-1/2}^{+} 
\Biggr \}
\end{multline}

To obtain moments for neutrino or antineutrino spectrum we have to add to \eqref{J2} the  term 
$\pm \Delta J_2$:
\begin{multline}
\label{deltaJ2}
\Delta J_2 =  \frac{ {G_F}^2 }{ 18 \pi } C_V C_A \;
\Bigl [ 
	4\, \left(  G_{3/2}^{-} G_{1/2}^{+} - G_{3/2}^{+} G_{1/2}^{-} \right)
	\\-
	\left(  G_{3/2}^{-} G_{-1/2}^{+} - G_{3/2}^{+} G_{-1/2}^{-} \right)
	+
	6\, \left(  G_{1}^{-} G_{0}^{+} - G_{1}^{+} G_{0}^{-} \right)
	\\+
	4\, \left(  G_{1/2}^{-} G_{-1/2}^{+} - G_{1/2}^{+} G_{-1/2}^{-} \right)
\Bigr ]
\end{multline}

Higher neutrino energy moments could be computed also, but very good approximation
for the whole spectrum can be obtained from the two first moments.

\subsection{Fitting formula \label{fit}}

Let's assume that the n-th moment $J_{1,2}^n$ of neutrino  spectrum $\lambda(\mathcal{E})$ is known, 
for example
from simulations, multidimensional integrations or numerical computations. Then, if we want to find 
parameters of the fitting formula $f$ with n parameters $\xi_i$:
\begin{equation}
\label{fit_general}
\lambda(\mathcal{E}) \simeq f(\mathcal{E}, \xi_1, \xi_2, \ldots \xi_n)
\end{equation}
we may require that our fitting
formula $f(\mathcal{E})$ has exactly the same moments as the original spectrum $\lambda(\mathcal{E})$,
given by e.g. \eqref{cuhre_widmo}:

\begin{equation}
\label{set}
J_i^n = \int_{0}^{\infty}  \mathcal{E}_i^n f(\mathcal{E}_i, \xi_1, \xi_2, \ldots \xi_n) \; d\mathcal{E}_i
= \int_0^\infty \mathcal{E}_i^n \lambda(\mathcal{E}_i) \; d\mathcal{E}_i
\end{equation}

Eqs.~\eqref{set} form a set of algebraic equations with unknown parameters $\xi_1,~\xi_2,\ldots\xi_n$.

Actually, only high energy tail of the neutrino spectrum can be used
to detect 
neutrinos\footnote{ 
Analytical description of the high-energy tail of the spectrum
could be very convenient, as the most of the existing big neutrino
detectors operate only for $\mathcal{E}>4$~MeV.
Authors, however, have failed to find analytical formulae for the 
tail of the spectrum.}. To find the high energy behavior of neutrinos produced in the process it is not enough
to know average $\nu$ energy. We must compute at least the second moment. We face immediately the problem 
of appropriate explicit form for
\eqref{fit_general}. For convenience we want to use as simple formula as possible. Very successful 
approximation of the neutrino spectrum produced in various situations is given by the following  
formula \cite{Mirizzi}:
\begin{equation}
\label{fitting_formula}
f(\mathcal{E}; \xi_1= \langle \mathcal{E} \rangle, \xi_2=\alpha) = \phi(\mathcal{E}) \equiv
\mathcal{N} 
\frac{\mathcal{E}^{\alpha}}{\langle \mathcal{E} \rangle^{\alpha+1}}
\exp\left( -\frac{\alpha+1 }{\langle \mathcal{E} \rangle}\; \mathcal{E} \right)
\end{equation}

Moments for this formula are:
\begin{equation}
\label{fit_moments}
\int_0^{\infty} \mathcal{E}^n \phi(\mathcal{E}) \; d\mathcal{E}
=
\mathcal{N}
\left( \frac{ \langle \mathcal{E} \rangle}{\alpha+1} \right)^n \frac{\Gamma(\alpha+1+n)}{\Gamma(\alpha + 1)}.
\end{equation}
Particularly, the normalization constant $\mathcal{N}$ is:
\begin{equation}
\mathcal{N} = \frac{(\alpha+1)^{\alpha+1}}{\Gamma(\alpha+1)},
\end{equation}
the first moment is, of course, equal to the average $\nu$ energy $\langle \mathcal{E} \rangle$:
\begin{equation}
\int_0^{\infty} \mathcal{E}\; \phi(\mathcal{E}) \; d\mathcal{E}= \langle \mathcal{E} \rangle
\end{equation}
and dispersion is given by the following expression:
\begin{equation}
\sigma_{\mathcal{E}} = \sqrt{\int_0^{\infty} (\mathcal{E} -  \langle \mathcal{E} \rangle)^2 \phi(\mathcal{E})  \; d\mathcal{E}} = \frac{ \langle \mathcal{E} \rangle}{\sqrt{\alpha+1}}
\end{equation}
Dispersion $\sigma_{\mathcal{E}}$ is a rough measure how far from the mean energy the spectrum extends.

If the spectrum is known, we can compute $\langle \mathcal{E} \rangle$ and $\sigma_{\mathcal{E}}$
which can be also computed
from \eqref{total_avg} and \eqref{J2}. One may also consider fitting the formula \eqref{fitting_formula}
to the spectrum in the least-squares sense if needed, but computing the moments is more straightforward.

Comparison of the spectrum and the fit \eqref{fitting_formula} is presented in Fig.~\ref{spectrum_vs_fit}.

Given the spectrum from Monte~Carlo simulations or computed using formula \eqref{cuhre_widmo},  it is easy 
to find parameters $\alpha$ and $\langle \mathcal{E} \rangle$
of the fitting formula \eqref{fitting_formula} for any value of the chemical potential
and temperature of the electron gas.

\section{Summary}

We have thoroughly analyzed details of the neutrino pair-annihilation process
in the electron-positron plasma. Plasma is assumed to be in thermal equilibrium defined by  
the temperature $T$ and the chemical potential $\mu$. Given previously known results, based on
the article of Dicus \cite{Dicus}, we are able to compute combined neutrino-antineutrino
emissivity $Q$ and the mean energy $\langle \mathcal{E} \rangle$. The latter however, was not presented
in the explicit form, so we derived the appropriate formula \eqref{total_avg}. Some useful analytical 
expressions
(\ref{R_ND}--\ref{D}) in various regimes are also presented for the neutrino energy. Further progress, 
namely derivation of separate neutrino and antineutrino emissivities and spectra is impossible with 
the use of Dicus cross-section \eqref{ovDicus}. Therefore we have derived cross-section \eqref{dsigma} for
pair annihilation into neutrino (antineutrino) of given energy in the plasma rest-frame. Next step, 
derivation of the spectrum becomes quite simple.
This spectrum is in full agreement (cf. Fig.~\ref{monte_vs_cuhre}) with results of the Monte~Carlo 
simulation.
Unfortunately, the value at any single point of the spectrum requires evaluation of the resulting 
(effectively three-dimensional) integral \eqref{cuhre_widmo}, which can only be evaluated numerically 
due to the presence of the unit step function \eqref{UnitStep} in the cross-section \eqref{dsigma}. 
Therefore, we provide formulae (\ref{Q},\ref{deltaQ},\ref{J2},\ref{deltaJ2}) for neutrino spectrum 
moments as a combination of the Fermi-Dirac integrals. Using just two first moments, we are able
to calculate parameters for the analytical approximation of the spectrum \eqref{fitting_formula}. 
In some cases, this procedure gives particularly simple expressions. For example, in relativistic 
and non-degenerate regime ($kT>2 \, m_e, \; kT>\mu$) the spectrum is given by:

\begin{equation}
\label{relativistic_fit}
\phi(\mathcal{E})= \frac{A}{kT}\; \left( \frac{\mathcal{E}}{kT} \right)^\alpha
\exp{(-a\,\mathcal{E}/{kT})}
\end{equation}

with parameters:
\begin{equation}
a = \frac{56700\; \zeta(5)\; \pi^4 }{ 217\; \pi^{10} - 13668750\; \zeta(5)^2}
\end{equation}
\begin{equation}
\alpha = \frac{217 \pi^{10} - 35538750 \; \zeta(5)^2 }{ 217 \pi^{10} - 13668750 \; \zeta(5)^2}
\end{equation}

\begin{equation}
A =  \frac{(\alpha+1)^{\alpha+1} }{\Gamma (\alpha+1)} \left ( \frac{2700 \; \zeta(5)}{7\; \pi^4} \right )^{-\alpha-1}
\end{equation}

\begin{figure}
\includegraphics[angle=270,width=8.6cm]{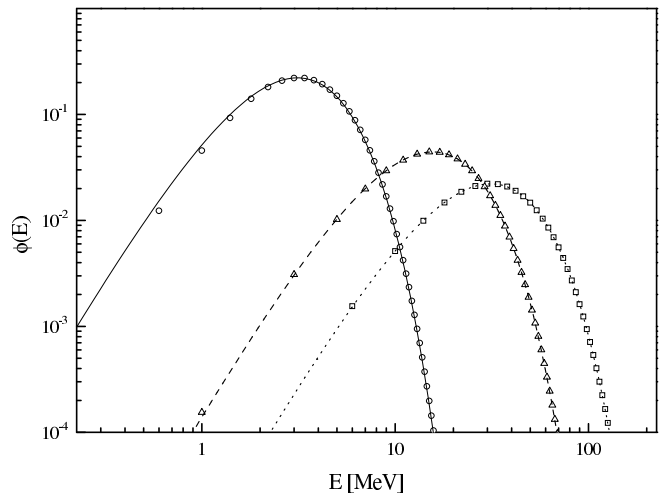}
\caption{\label{spectrum_vs_fit} Numerically computed points of the neutrino spectrum (symbols) and the fit (lines). Curves represent spectra for $kT=1, 5, 10$~MeV.  Even in in the worst case considered, $kT=1$~MeV ($\circ$),  fit  \eqref{relativistic_fit} is particularly good. }
\end{figure}

Numerically: \mbox{$\alpha=3.180657028$}, \mbox{$A=0.1425776426$}, \mbox{$a=1.018192299$}.
Previously, the spectrum given by the analytical formula \eqref{relativistic_fit} above, must have 
been computed by the means of the Monte~Carlo simulation! Comparison of the spectrum computed numerically
and given by eq.~\eqref{relativistic_fit} is presented in Fig.~\ref{spectrum_vs_fit}.

In more difficult cases, e.g. required for neutrinos from the pre-supernova core, where 
$kT\sim m_e \sim \mu$, one must use general expressions for neutrino energy moments, 
with wealth of analytical \cite{BlinnikovRudzskij, FDanal1} and numerical \cite{FDnum1, FDnum2} methods 
available for calculating Fermi-Dirac integrals. In the degenerate case
differences between neutrino and antineutrino spectrum (and between $\nu_e$ and $\nu_{\mu, \tau}$ as well,
cf. Fig.~\ref{4_flavours}) must be taken into account.
However, in typical situations they may be considered as a small perturbation 
to the average value.

If for some reasons exact results are required, the spectrum may be computed from \eqref{cuhre_widmo} 
point-by-point as well.

\acknowledgments
This work was supported by grant of Polish  Ministry of Education and Science (former
Ministry of Scientific Research and Information Technology, 
now Ministry of Science and Higher Education) No. 1~P03D~005~28.
M.~Misiaszek  was partly supported by EU Marie Curie Fellowship
HPMT-CT-2001-00279.

\bibliography{Pair-annihilation}

\end{document}